\documentclass[12pt]{article}
\pdfoutput=1

\usepackage{color}
\usepackage{graphicx}
\usepackage{amsmath}
\usepackage{slashed}
\usepackage{amssymb}
\usepackage{epstopdf}
\usepackage{bbold}

\input xy
\xyoption{all}
\xyoption{web}

\usepackage[
      colorlinks=true,
      linkcolor=blue,
      urlcolor=blue,
      filecolor=black,
      citecolor=red,
      pdfstartview=FitV,
      pdftitle={},
        pdfauthor={Yi Li},
        pdfsubject={},
        pdfkeywords={},
        pdfpagemode={},
        bookmarksopen=true
      ]{hyperref}

\usepackage{color}
\usepackage{graphicx}

\usepackage{sectsty}
\sectionfont{\large}

\renewcommand{\thefootnote}{\fnsymbol{footnote}}
\renewcommand{\thanks}[1]{\footnote{#1}}
\newcommand{\starttext}{
\setcounter{footnote}{0}
\renewcommand{\thefootnote}{\arabic{footnote}}}

\newcommand{\bea}{\begin{eqnarray}}
\newcommand{\eea}{\end{eqnarray}}
\newcommand{\be}{\begin{equation}}
\newcommand{\ee}{\end{equation}}

\renewcommand{\>}{\rangle}

\def\det{{\rm det}}

\usepackage{ dsfont } 

\long\def\symbolfootnote[#1]#2{\begingroup%
\def\thefootnote{\fnsymbol{footnote}}\footnote[#1]{#2}\endgroup}

\begin{document}
\setlength{\baselineskip}{18pt}

\starttext
\setcounter{footnote}{0}

%
\bigskip

\begin{center}

{\Large \bf  Comments on large central charge $T\bar{T}$ deformed conformal field theory and cutoff AdS holography}

\vskip 0.4in

{\large Yi Li}

\vskip .2in

{\it Department of Physics and Center for Field Theory and Particle Physics}\\
{\it Fudan University, Shanghai 200438, China}\\[0.5cm]
\href{mailto:liyi@fudan.edu.cn}{\texttt{liyi@fudan.edu.cn}}

\bigskip

\bigskip

\end{center}

\begin{abstract}

\setlength{\baselineskip}{18pt}

In this article we study large central charge partition function and entanglement entropy of $T\bar{T}$ deformed two dimensional conformal field theory, following the approach to $T\bar{T}$ deformation as integrated infinitesimal double trace deformation used by Guica et al.. For sphere partition function and entanglement entropy of half great circle with antipodal points being the entangling surface, we obtain different results compared to previous works, with more reasonable CFT limits and qualitatively different behaviour as the deformation parameter $\mu$ goes to infinity, which contradicts the simple version of the cutoff AdS holography proposal. For a modified version of cutoff AdS holography which is supposed to work only in the sector of classical pure gravity, we show that the flow equation of the metric and one point function of energy-momentum tensor in $T\bar{T}$ deformation corresponds to the flow equation of the boundary metric and Brown-York tensor on a cutoff surface in AdS space as the cutoff surface moves in the direction of normal geodesics. In addition the flow equation of gravity on-shell action takes the form of $T\bar{T}$ deformation, with straightforward generalization to higher dimensions. As an example we give a holographic computation of the sphere partition function of $T\bar{T}$ CFT.
\end{abstract}

\setcounter{equation}{0}
\setcounter{footnote}{0}

%
%
%
%
%
\newpage


\section{Introduction}
\setcounter{equation}{0}
\label{sec1}

For a two dimensional quantum field theory with a symmetric and conserved energy-momentum tensor $T_{ij}$ \footnote{We use the convention for the energy-momentum tensor $\delta S = \frac{1}{2}\int d^dx\sqrt{\gamma} T_{ij}\delta \gamma^{ij}$.}, the composite $T\bar{T}$ operator
\begin{align}
 T\bar{T} = T^{ij}T_{ij} - T^2
\end{align}
can be universally defined in flat space with many interesting properties \cite{Zamolodchikov:2004ce}. Deformation by the operator, known as $T\bar{T}$ deformation, is defined as
\begin{align}
 \frac{d S^{[\mu]}}{d\mu} = \frac{1}{8}\int d^dx\sqrt{\gamma} T\bar{T}^{[\mu]}
\end{align}
where the $T\bar{T}$ operator varies together with the action since the energy-momentum tensor itself depends on the action. Finite $T\bar{T}$ deformation changes the theory drastically, in general makes it non-local but well-behaved in many ways \cite{Smirnov:2016lqw}\cite{Cavaglia:2016oda}\cite{Dubovsky:2012wk}\cite{Dubovsky:2017cnj}\cite{Dubovsky:2018bmo}\cite{Frolov:2019nrr}\cite{Sfondrini:2019smd}\cite{Callebaut:2019omt}\cite{Jorjadze:2020ili}. It has solvable finite size spectrum \cite{Zamolodchikov:2004ce}\cite{Smirnov:2016lqw}\cite{Cavaglia:2016oda}, deforms the scattering matrix by multiplying CDD factors \cite{Cavaglia:2016oda}\cite{Dubovsky:2017cnj}, and preserves integrability structures, symmetries and modular properties \cite{Smirnov:2016lqw}\cite{Conti:2018jho}\cite{LeFloch:2019wlf}\cite{Chen:2021aid}\cite{Guica:2020uhm}\cite{Guica:2020eab}\cite{Datta:2018thy}\cite{Aharony:2018bad}. $T\bar{T}$ deformation has been intensively studied in the past few years. Explicit form of $T\bar{T}$ deformed Lagrangian was computed for several models in \cite{Bonelli:2018kik}. A illuminating perturbative analysis was given in \cite{Rosenhaus:2019utc}. The partition function and its flow equation were studied in \cite{Dubovsky:2018bmo}\cite{Datta:2018thy}\cite{Cardy:2018sdv}\cite{Donnelly:2018bef}\cite{Santilli:2018xux}\cite{Caputa:2019pam}\cite{He:2020cxp}. Correlation functions of $T\bar{T}$ deformed QFT or CFT were studied in \cite{Kraus:2018xrn}\cite{Aharony:2018vux}\cite{Cardy:2019qao}\cite{He:2019vzf}\cite{He:2019ahx}\cite{He:2020udl}\cite{Li:2020pwa}\cite{Hirano:2020ppu}\cite{He:2020qcs}. $T\bar{T}$ deformation of supersymmetric theories was considered in \cite{He:2019ahx}\cite{Baggio:2018rpv}\cite{Chang:2018dge}\cite{Jiang:2019hux}\cite{Chang:2019kiu}. While much of the work on the $T\bar{T}$ deformation has been done in the flat space, generalization to curved spaces was considered in \cite{Jiang:2019tcq}\cite{Brennan:2020dkw}\cite{Tolley:2019nmm}\cite{Mazenc:2019cfg}. Quantum entanglement in $T\bar{T}$ CFT was studied in \cite{Donnelly:2018bef}\cite{Chakraborty:2018kpr}\cite{Chen:2018eqk}\cite{Banerjee:2019ewu}\cite{Murdia:2019fax}\cite{Ota:2019yfe}\cite{Jeong:2019ylz}\cite{Grieninger:2019zts}\cite{Lewkowycz:2019xse}\cite{Geng:2019ruz}\cite{Donnelly:2019pie}\cite{Chakraborty:2020udr}\cite{Khoeini-Moghaddam:2020ymm} with strong connection to holography. Other developments in $T\bar{T}$ deformation include but are not limited to \cite{Conti:2018tca}\cite{Jafari:2019qns}\cite{Gross:2019ach}\cite{Ireland:2019vvj}\cite{Haruna:2020wjw}\cite{Leoni:2020rof}\cite{Kruthoff:2020hsi}\cite{Ouyang:2020rpq}\cite{Santilli:2020qvd}\cite{Jiang:2020nnb}\cite{He:2020hhm}\cite{Hansen:2020hrs}.

Considering these interesting properties of $T\bar{T}$ deformation, it's tempting to ask what its holographic dual is for a holographic CFT. Among other proposals \cite{Giveon:2017myj}\cite{Hirano:2020nwq}, it was proposed by McGough et al. that for positive $T\bar{T}$ deformation parameter $\mu$ the holographic dual of $T\bar{T}$ deformation is a Dirichlet cutoff in the $\rm AdS_3$ gravity, based on computation of signal propagation speed, quasi-local energy of BTZ blackhole and other physics quantities \cite{McGough:2016lol}. This proposal was supported by many subsequent studies, mostly working in the Einstein's gravity sector  which we briefly review here. The Euclidean gravity action for ${\rm AdS}_{d+1}$ space is
\begin{align}
 I_{grav} &= I^{BY} + I^{CT} \nonumber\\
 &= -\frac{1}{16 \pi G}\int_{\cal B} d^{d+1}x \sqrt{g} (R + \frac{d(d-1)}{l^2}) - \frac{1}{8\pi G} \int_{{\cal M}=\partial{\cal B}} d^dy \sqrt{h} K + \frac{1}{8\pi G} \int_{{\cal M}=\partial{\cal B}}  d^dy \sqrt{h} \frac{d-1}{l} + I^{CT}
\end{align}
with $I^{CT}$ consisted of local counter terms constructed from the boundary metric $h_{ij}$, and its form depends on the dimension $d$ \cite{Balasubramanian:1999re}\cite{Henningson:1998gx}\cite{deHaro:2000vlm}. The holographic energy-momentum tensor is
\begin{align}
 {T_{grav}}_{ij} &= T^{BY}_{ij} + T^{CT}_{ij} \nonumber\\
 &=-\frac{1}{8\pi G} (K_{ij} - K h_{ij} + \frac{d-1}{l}h_{ij}) + T^{CT}_{ij}
\end{align}
where for simplicity of discussion we define the Brown-York tensor $T^{BY}_{ij}$ to include the term $\frac{d-1}{l}h_{ij}$. For $d=2$, $I^{CT} = -\frac{1}{32\pi G}\log\epsilon \int_{\cal M} d^2x \sqrt{h} \hat{R}(h)$ is a Weyl anomaly term depending on the IR cutoff $\epsilon$. In addition $\int_{\cal M} d^2x \sqrt{h} \hat{R}(h)=4\pi \chi(\cal M)$ is a topological invariant where $\chi(\cal M)$ is the Euler characteristic. So $I^{CT}$ doesn't contribute to the holographic energy-momentum tensor for $d=2$, we simply have ${T_{grav}}_{ij} = T^{BY}_{ij}$. The simple version of cutoff AdS holography puts the $T\bar{T}$ CFT on a finite cutoff surface in the AdS space, usually a surface of constant radial Fefferman-Graham coordinate $\rho_c$, with the induced metric as the background metric for the field theory, and identifies the on-shell dynamics of classical gravity inside the cutoff as the large $c$ dynamics of the $T\bar{T}$ CFT with the following dictionary between the gravity parameters and the $T\bar{T}$ CFT parameters
\begin{align} \label{OldDictionary}
 c=\frac{3l}{2G}, \quad \mu = 16\pi G l
\end{align}
Correlators of energy-momentum tensor were computed in \cite{Kraus:2018xrn}\cite{Li:2020pwa}, with heavy use of the trace relation in the field theory computation
\begin{align} \label{TraceRelationMu}
 T = \frac{\mu}{4} T\bar{T} + \frac{c}{24\pi}R(\gamma)
\end{align}
The trace relation was identified as one of the Gauss-Codazzi equations \footnote{Or one component of the Einstein's equation.} of the Brown-York tensor $T^{BY}_{ij}$ \cite{Kraus:2018xrn}\cite{Taylor:2018xcy}, which suggests a generalization of $T\bar{T}$ operator to a higher dimension $d$ \cite{Taylor:2018xcy}
\begin{align} \label{TTbarGeneralDimension}
 T\bar{T} = T_{ij}T^{ij} - \frac{1}{d-1}T^2
\end{align} 
In addition, sphere partition function and entanglement entropy of half great circle were computed in \cite{Donnelly:2018bef} as strong support to cutoff AdS holography. The partition function was found to agree with the gravity on-shell action of cutoff AdS and the entanglement entropy was found to agree with the minimal area prescription of Ryu-Takayanagi formula \cite{Ryu:2006bv}\cite{Lewkowycz:2013nqa}, though it's a bit concerning to us that the results do not have a well-defined CFT limit. Higher dimensional generalization was subsequently studied in \cite{Caputa:2019pam}\cite{Banerjee:2019ewu}. The relation between $T\bar{T}$ CFT and cutoff AdS was also studied in \cite{Hartman:2018tkw}, where the metric of the field theory was identified with induced metric multiplied by the radial Fefferman-Graham coordinate at the cutoff position $\rho_c$ following the spirit of AdS/CFT. That should result in a change of the dictionary between the gravity parameters and the $T\bar{T}$ CFT parameters, recently shown in \cite{Caputa:2020lpa}
\begin{align} \label{NewDictionary}
 c=\frac{3l}{2G}, \quad \mu = 16\pi G l \rho_c
\end{align}
It was also shown in \cite{Caputa:2020lpa} that the gravity on-shell action in the annular region between two cutoff surfaces is proportional to the integral of $T\bar{T}^{BY} = T^{BY}_{ij}{T^{BY}}^{ij} - {T^{BY}}^2$ over either of the cutoff surfaces. Other developments in the relation between $T\bar{T}$ CFT and cutoff AdS, with generalization to other dimensions, include \cite{Iliesiu:2020zld}\cite{Belin:2020oib}. Of particular interests to us, an important work in the relation between $T\bar{T}$ CFT and cutoff AdS was \cite{Guica:2019nzm} which derived the $T\bar{T}$ flow equation of the generating functional, the metric and the one point function of the energy-momentum tensor in the large $c$ limit. It showed that the trace relation follows from the flow equation, and more importantly, the one point function as a functional of the metric holographically corresponds to the Brown-York tensor as a functional of the boundary metric on the cutoff surface. It explained most previous evidences supporting cutoff AdS holography, those involved physics quantities that can be computed from the one point function. Remarks on the limitation of the cutoff AdS holography for $T\bar{T}$ CFT were also made in \cite{Guica:2019nzm}, especially when matter fields are added to pure AdS gravity.

With primary focus on cutoff AdS holography, we organize our article as follows. In Section \ref{sec2} we review the derivation of the $T\bar{T}$ flow equation in the large $c$ limit in \cite{Guica:2019nzm}, hoping to make it a bit more accessible. In section \ref{sec3} we use the $T\bar{T}$ flow equation to compute the sphere partition function and the entanglement entropy of half great circle with antipodal points being the entangling surface, with comparison to previous work in \cite{Donnelly:2018bef}. In section \ref{sec4} we derive the cutoff AdS holography in the classical pure gravity sector by identifying the flow equation of the metric and one point function with the flow equation of the boundary metric and Brown-York tensor on a cutoff surface in AdS space as it moves inward. We also show the flow equation of the on-shell gravity action takes the form of $T\bar{T}$ deformation, in two and higher dimensions. Therefore we take a slightly different viewpoint than that in \cite{Guica:2019nzm}, as far as classical pure gravity is concerned we think the Dirichlet cutoff is not a ``mirage", and the mixed boundary condition involving the boundary metric and finite number of its normal derivatives is what the Dirichlet cutoff happens to take the form of in three dimensional gravity. We end section \ref{sec4} with a holographic computation of sphere partition function and complete the article with summary and discussion in section \ref{sec5}.

\section{Solving the $T\bar{T}$ deformation in the large central charge limit}
\setcounter{equation}{0}
\label{sec2}
In this section we briefly review the approach to $T\bar{T}$ deformed conformal field theory in the large central charge limit used by Guica et al.\cite{Guica:2019nzm}. The large central charge limit is the semi-classical limit of conformal field theory of large degrees of freedom. It corresponds to the large $N$ limit for holographic gauge theories. $T\bar{T}$ deformation can be taken as composition of successive infinitesimal double trace deformations, with the double trace operator $T\bar{T}$ changes as the deformation goes. Double trace deformation, and more generally multi-trace deformation, have been studied in \cite{Witten:2001ua}\cite{Berkooz:2002ug}\cite{Mueck:2002gm}\cite{Minces:2002wp}\cite{Sever:2002fk}\cite{Gubser:2002zh}\cite{Gubser:2002vv}\cite{Aharony:2005sh}\cite{Elitzur:2005kz}\cite{Hartman:2006dy}\cite{Diaz:2007an}\cite{Papadimitriou:2007sj}\cite{Giombi:2018vtc}, in particular for its holographic dual as changing boundary conditions in AdS space. We start with a generic example of double trace deformation and then apply the method to $T\bar{T}$ deformation.

Consider a theory defined by the action $S[\phi]$ with $\phi$ being the fundamental field in path integration, and a single trace operator $O$. The generating functional for the operator $O$ is defined as
\begin{align}
 I^{[0]}[J] = -\log Z^{[0]}[J]
\end{align}
where
\begin{align}
 Z^{[0]}[J] = \int {\cal D}\phi e^{-S[\phi]+\int d^dx \sqrt{\gamma} J O}
\end{align}
Connected correlation functions of the operator $O$ are given by the functional derivative of the generating functional $I^{[0]}[J]$ with respect to the source $J$, and in particular for one point function
\begin{align}
 \<O(x)\>^{[0]}_J = -\frac{1}{\sqrt{\gamma(x)}}\frac{\delta I^{[0]}[J]}{\delta J(x)}
\end{align}
Suppose the theory is deformed by the double trace deformation
\begin{align}
 S[\phi] \to S[\phi] + \frac{f}{2} \int d^dx\sqrt{\gamma} O^2
\end{align}
where the constant $f$ is the deformation parameter. The deformed partition function with source $J$ is
\begin{align} \label{DoubleTraceDeformationPartitionFunction}
 Z^{[f]}[J] = \int {\cal D}\phi e^{-S[\phi]+\int d^dx \sqrt{\gamma} J O - \frac{f}{2}\int d^dx \sqrt{\gamma}O^2}
\end{align}
By a Hubbard-Stratonovich transformation, that is, inserting the identity into the path integral
\begin{align}
 1 = \sqrt{\det(-\frac{1}{f}\mathbb{1})}\int {\cal D}\sigma e^{\frac{1}{2f}\int d^dx \sqrt{\gamma} (\sigma + f O)^2}
\end{align}
where $\mathbb{1}$ is schematically the operator that takes two functions $g_1, g_2$ to $\int d^dx \sqrt{\gamma} g_1 g_2$, we find
\begin{align}
 Z^{[f]}[J] &= \int {\cal D}\phi \sqrt{\det(-\frac{1}{f}\mathbb{1})}\int {\cal D}\sigma e^{-S[\phi]+\int d^dx \sqrt{\gamma} (J+\sigma) O + \frac{1}{2f}\int d^dx \sqrt{\gamma} \sigma^2} \nonumber\\
 & = \sqrt{\det(-\frac{1}{f}\mathbb{1})}\int {\cal D}\sigma e^{\frac{1}{2f}\int d^dx \sqrt{\gamma} \sigma^2} \int {\cal D}\phi e^{-S[\phi]+\int d^dx \sqrt{\gamma} (J+\sigma) O} \nonumber\\
 & = \sqrt{\det(-\frac{1}{f}\mathbb{1})}\int {\cal D}\sigma e^{\frac{1}{2f}\int d^dx \sqrt{\gamma} \sigma^2 - I^{[0]}[J+\sigma]}
\end{align}
Shifting the variable of path integration $\sigma^{'} = \sigma + J$, we get
\begin{align}
 Z^{[f]}[J] = \sqrt{\det(-\frac{1}{f}\mathbb{1})}\int {\cal D}\sigma^{'} e^{\frac{1}{2f}\int d^dx \sqrt{\gamma} (\sigma^{'}-J)^2 - I^{[0]}[\sigma^{'}]}
\end{align}
The equation for the saddle point $\sigma^*$ is
\begin{align}
 \frac{1}{f} (\sigma^{'}-J) - \frac{1}{\sqrt{\gamma}} \frac{\delta I^{[0]}[\sigma^{'}]}{\delta \sigma^{'}} = 0
\end{align}
that is
\begin{align}
 J = \sigma^* - f \frac{1}{\sqrt{\gamma}} \frac{\delta I^{[0]}[\sigma^{'}]}{\delta \sigma^{'}}|_{\sigma^*} = \sigma^* + f \<O\>^{[0]}_{\sigma^*}
\end{align}
Expanding around the saddle point and substituting in $\xi=\sigma^{'}-\sigma^*$ we get
\begin{align}
 Z^{[f]}[J] =& e^{-I^{[0]}[\sigma^*] + \frac{f}{2}\int d^dx \sqrt{\gamma} (\<O\>^{[0]}_{\sigma^*})^2} \sqrt{\det(-\frac{1}{f}\mathbb{1})}\int {\cal D}\xi e^{\frac{1}{2f}\int d^dx\sqrt{\gamma} \xi^2 - \frac{1}{2!}\int d^dx  \int d^dy  \frac{\delta^2 I^{[0]}[\sigma^{'}]}{\delta \sigma^{'}(x) \delta \sigma^{'}(y)}|_{\sigma^*} \xi(x)\xi(y) + \ldots} \nonumber\\
 =& e^{-I^{[0]}[\sigma^*] + \frac{f}{2}\int d^dx \sqrt{\gamma} (\<O\>^{[0]}_{\sigma^*})^2} \sqrt{\det(-\mathbb{1})}\int {\cal D}\zeta e^{\frac{1}{2}\int d^dx\sqrt{\gamma} \zeta^2 - f \frac{1}{2!}\int d^dx \sqrt{\gamma(x)} \int d^dy \sqrt{\gamma(y)} \<O(x)O(y)\>^{[0]c}_{\sigma^*} \zeta(x)\zeta(y) + \ldots} \nonumber\\
\end{align}
where in the second line we did a change of variable $\zeta^2 = \frac{1}{f}\xi^2$. The omitted part in the expression is the contribution from connected higher point functions with higher powers in $f$. For an infinitesimal deformation parameter $f$, only the term of connected two point function $\<O(x)O(y)\>^{[0]c}_{\sigma^*}$ survives to the first order in $f$, but it's dominated by the product of two one point functions in front of the path integral in the large $c$ limit. So in the large c limit we have the saddle point approximation
\begin{align}
 Z^{[f]}[J] = e^{-I^{[0]}[\sigma^*] + \frac{f}{2}\int d^dx \sqrt{\gamma} (\<O\>^{[0]}_{\sigma^*})^2}
\end{align}
or by rebranding $\sigma^*$ as the "undeformed source" $J^{[0]}$
\begin{align}
 I^{[f]}[J] = I^{[0]}[J^{[0]}] - \frac{f}{2}\int d^dx \sqrt{\gamma} (\<O\>^{[0]}_{J^{[0]}})^2
\end{align}
with $J = J^{[0]} + f \<O\>^{[0]}_{J^{[0]}}$. Regarding a finite double trace deformation as a composition of successive infinitesimal ones with the deformation parameter $f$ varying continuously, we have the flow equation for the generating functional and the source
\begin{align}
 \partial_f I^{[f]}[J^{[f]}] &= -\frac{1}{2} \int d^dx \sqrt{\gamma} (\<O\>^{[f]}_{J^{[f]}})^2 \nonumber\\
 \partial_f J^{[f]} &= \<O\>^{[f]}_{J^{[f]}}
\end{align}
Using the chain rule we have
\begin{align}
 \partial_f I^{[f]}[J^{[f]}] = (\partial_f I^{[f]})[J^{[f]}] + \int d^dx \frac{\delta I^{[f]}[J^{[f]}]}{\delta J^{[f]}} \partial_f J^{[f]} = (\partial_f I^{[f]})[J^{[f]}] - \int d^dx \sqrt{\gamma} (\<O\>^{[f]}_{J^{[f]}})^2
\end{align}
So we can fix the source and single out the flow of the generating functional
\begin{align}
 (\partial_f I^{[f]})[J] = \frac{1}{2} \int d^dx \sqrt{\gamma} (\<O\>^{[f]}_J)^2
\end{align}
This is equivalent to replacing $O$ by its one point function in the double trace operator in (\ref{DoubleTraceDeformationPartitionFunction}).

Now we apply this method to the $T\bar{T}$ deformation, mainly for $\rm CFT_2$ but the derivation doesn't really depends on the dimension. The source is the background metric, the conjugate operator is the energy-momentum tensor and the double trace operator is the $T\bar{T}$ operator. For an infinitesimal step as the deformation parameter $\lambda$ goes to $\lambda+\Delta\lambda$, the partition function becomes
\begin{align} \label{InfTTbarPathIntegral}
 Z^{'}[\gamma^{'}_{ij}] = \int {\cal D}\phi e^{-S[\phi,\gamma^{'}_{ij}] - \Delta\lambda S_{T\bar{T}}}
\end{align}
with the $T\bar{T}$ deformation action being
\begin{align}
 S_{T\bar{T}} = -\frac{1}{2}\int d^2x \sqrt{\gamma^{'}} T\bar{T} = -\frac{1}{2}\int d^2x \sqrt{\gamma^{'}} (T_{ij}T^{ij} - T^2) = \frac{1}{2} \int d^2 x \sqrt{\gamma^{'}} \epsilon^{ik}\epsilon^{jl}T_{ij}T_{kl}
\end{align}
Here $\lambda$ is the $T\bar{T}$ deformation parameter defined in \cite{Guica:2019nzm}, denoted by $\mu$ there. It's related to $\mu$ used in our work and other references \cite{Donnelly:2018bef}\cite{McGough:2016lol} by
\begin{align}
 \lambda = -\frac{\mu}{4}
\end{align}
By doing a Hubbard-Stratonovich transformation \footnote{This transformation is also used in \cite{Cardy:2018sdv} to derive the flow equation of partition functions in spaces with simple geometries where considerable simplification of the path integral over $h$ is possible.}, that is, inserting in the identity
\begin{align}
 1 = \sqrt{\det(-\Delta\lambda M_{\gamma^{'}_{ij}})} \int {\cal D}h e^{\frac{1}{2}\Delta\lambda \int d^2x\sqrt{\gamma^{'}} \epsilon^{ik}\epsilon^{ij}(h_{ij}+T_{ij})(h_{kl}+T_{kl})}
\end{align}
where $M_{\gamma^{'}_{ij}}$ is schematically the operator $M_{\gamma^{'}_{ij}}(f_{kl},g_{mn}) = \int d^2x\sqrt{\gamma^{'}} \epsilon^{km}\epsilon^{ln} f_{kl} g_{mn}$, we find
\begin{align}
 Z^{'}[\gamma^{'}_{ij}] &= \int {\cal D}\phi \sqrt{\det(-\Delta\lambda M_{\gamma^{'}_{ij}})} \int {\cal D}h e^{-S[\phi,\gamma^{'}_{ij}] + \frac{1}{2}\Delta\lambda \int d^2x \sqrt{\gamma^{'}} \epsilon^{ik}\epsilon^{jl}h_{ij}h_{kl} + \Delta\lambda \int d^2x \sqrt{\gamma^{'}} \epsilon^{ik}\epsilon^{jl}T_{ij}h_{kl}} \nonumber\\
 &=\sqrt{\det(-\Delta\lambda M_{\gamma^{'}_{ij}})} \int {\cal D}h e^{\frac{1}{2}\Delta\lambda \int d^2x \sqrt{\gamma^{'}} \epsilon^{ik}\epsilon^{jl}h_{ij}h_{kl}} \int {\cal D}\phi  e^{-S[\phi,\gamma^{'}_{ij}]  + \Delta\lambda \int d^2x \sqrt{\gamma^{'}} \epsilon^{ik}\epsilon^{jl}T_{ij}h_{kl}} \nonumber\\
 &\approx\sqrt{\det(-\Delta\lambda M_{\gamma^{'}_{ij}})} \int {\cal D}h e^{\frac{1}{2}\Delta\lambda \int d^2x \sqrt{\gamma^{'}} \epsilon^{ik}\epsilon^{jl}h_{ij}h_{kl}} \int {\cal D}\phi  e^{-S[\phi,{\gamma^{'}}^{ij} - 2\Delta\lambda \epsilon^{ik}\epsilon^{jl}h_{kl}]} \nonumber\\
 &=\sqrt{\det(-\Delta\lambda M_{\gamma^{'}_{ij}})} \int {\cal D}h e^{\frac{1}{2}\Delta\lambda \int d^2x \sqrt{\gamma^{'}} \epsilon^{ik}\epsilon^{jl}h_{ij}h_{kl} - I[{\gamma^{'}}^{ij} - 2\Delta\lambda \epsilon^{ik}\epsilon^{jl}h_{kl}]}
\end{align}
where in the third line we neglect terms of high order in $\Delta \lambda$. In the large $c$ limit we find the saddle point of the path integral by
\begin{align}
 \Delta\lambda \epsilon^{ik} \epsilon^{jl} h_{ij} - \frac{1}{\sqrt{\gamma}} \frac{\delta I}{\delta\gamma^{ij}} (-2\Delta\lambda \epsilon^{ik}\epsilon^{jl}) = 0
\end{align}
So we have the saddle point $h_{ij}^* = -\<T_{ij}\>$, $\<T_{ij}\>$ is the one point function of the energy-momentum tensor, as a functional of the metric $\gamma^{ij} = {\gamma^{'}}^{ij}+ 2\Delta\lambda \epsilon^{ik}\epsilon^{jl}\<T\>_{ij}$ by the relation in the undeformed theory $\delta I[\gamma_{ij}] = \frac{1}{2}\int d^2x \sqrt{\gamma} \<T\>_{ij} \delta\gamma^{ij}$. Using the saddle point approximation we find to first order in $\Delta\lambda$
\begin{align}
 I^{'}[\gamma^{'}_{ij}] = I[{\gamma_{ij}}] - \frac{1}{2} \Delta\lambda \int d^2x \sqrt{\gamma} \epsilon^{ik}\epsilon^{jl}\<T_{ij}\>\<T_{kl}\>
\end{align}
Taking the limit $\Delta \lambda \to 0$ we have
\begin{align} \label{TTbarflowGeneratingFunctional}
 \partial_\lambda I^{[\lambda]}[\gamma^{[\lambda]}_{ij}] = \frac{1}{2}\int d^2x \sqrt{\gamma^{[\lambda]}} T\bar{T}^{[\lambda]}
\end{align}
with
\begin{align} \label{TTbarflowMetric}
 \partial_\lambda {\gamma^{[\lambda]}}^{ij} = -2{\epsilon^{[\lambda]}}^{ik}{\epsilon^{[\lambda]}}^{jl} T^{[\lambda]}_{kl} = 2({T^{[\lambda]}}^{ij} - T^{[\lambda]} {\gamma^{[\lambda]}}^{ij})
\end{align}
Here we denote the one point function of energy-momentum tensor just by $T_{ij}$ for simplicity, and we stick to this convention from now on. Similar to the previous case of a generic double trace deformation, we can single out the flow of the generating functional itself with metric fixed
\begin{align} \label{TTbarflowGeneratingFunctionalMetricFixed}
 (\partial_\lambda I^{[\lambda]})[\gamma] = -\frac{1}{2} \int d^2x \sqrt{\gamma} T\bar{T}^{[\lambda]}
\end{align}
where $T^{[\lambda]}_{ij}$ is a functional of the metric that depends on $I^{[\lambda]}$ as a functional of the metric. This is what we get if we replace the energy-momentum tensor by its one point function in the $T\bar{T}$ operator in the path integral in (\ref{InfTTbarPathIntegral}). In general (\ref{TTbarflowGeneratingFunctionalMetricFixed}) is a complicated functional equation for the generating functional. For explicit flow equation of the one point function, we take a variation in the metric  of (\ref{TTbarflowGeneratingFunctional}) and obtain
\begin{align} \label{VariationalPrinciple}
 \partial_\lambda ( \sqrt{\gamma^{[\lambda]}}T^{[\lambda]}_{ij} \delta {\gamma^{[\lambda]}}^{ij}) = \delta(\sqrt{\gamma^{[\lambda]}}T\bar{T}^{[\lambda]})
\end{align}
which is dubbed the variational principle in \cite{Guica:2019nzm}\cite{Bzowski:2018pcy}. It was argued in \cite{Guica:2019nzm}\cite{Bzowski:2018pcy} that similar formula holds for generating functional with sources dual to other operators turned on
\begin{align} \label{TTbarflowSourcesAndOperators}
 \partial_\lambda (\frac{1}{2} \sqrt{\gamma^{[\lambda]}}T^{[\lambda]}_{ij} \delta {\gamma^{[\lambda]}}^{ij} + \sum_A \sqrt{\gamma^{[\lambda]}} O_A^{[\lambda]} \delta{J^A}^{[\lambda]}) = \frac{1}{2} \delta(\sqrt{\gamma^{[\lambda]}}T\bar{T}^{[\lambda]})
\end{align}
We restrict ourselves to the case without other sources. The equation (\ref{VariationalPrinciple}) was rewritten in \cite{Guica:2019nzm} as
\begin{align} \label{TTbarfromVariationalPrinciple}
 \partial_\lambda (\sqrt{\gamma}T_{ij}) \delta\gamma^{ij}  + \sqrt{\gamma}T_{ij} \delta(\partial_\lambda \gamma^{ij}) = \sqrt{\gamma}[(-\frac{1}{2}T\bar{T}\gamma_{ij}-2T_{ik}T_j^k+2TT_{ij})\delta\gamma^{ij} + T_{ij}\delta(2T^{ij}-2\gamma^{ij}T)]
\end{align}
and the following $T\bar{T}$ flow equations for the metric and one point function were obtained
\begin{align}
 &\partial_\lambda \gamma^{ij} = 2(T^{ij}-\gamma^{ij}T) \nonumber\\
 &\partial_\lambda(\sqrt{\gamma}T_{ij}) = \sqrt{\gamma}(-\frac{1}{2}T\bar{T}\gamma_{ij}-2T_{ik}T_j^k+2TT_{ij})
\end{align}
where we have omitted the superscript $[\lambda]$ with the understanding that the metric and one point function depend on $\lambda$.
To derive it from (\ref{TTbarfromVariationalPrinciple}) we have used (\ref{TTbarflowMetric}) then the other equation follows.  It was further reduced to a simpler form by defining $\hat{T}_{ij} = T_{ij} - T\gamma_{ij}$
\begin{align} \label{TTbarflowMetricEMT2dim}
 &\partial_\lambda \gamma_{ij} = -2 \hat{T}_{ij} \nonumber\\
 &\partial_\lambda \hat{T}_{ij} = \frac{1}{2} T\bar{T}\gamma_{ij} - 2\hat{T}_{ik}\hat{T}_j^k + \hat{T}\hat{T}_{ij}
\end{align}
The solution to these equations was  found to be \cite{Guica:2019nzm}
\begin{align} \label{TTbarflowSol2dim}
 \gamma^{[\lambda]}_{ij} &= \gamma^{[0]} _{ij}- 2\lambda \hat{T}^{[0]}_{ij} + \lambda^2 \hat{T}^{[0]}_{ik} \gamma^{[0]kl} \hat{T}^{[0]}_{lj} \nonumber\\
 \hat{T}^{[\lambda]}_{ij} &= \hat{T}^{[0]}_{ij} - \lambda \hat{T}^{[0]}_{ik} \gamma^{[0]kl} \hat{T}^{[0]}_{lj}
\end{align}
It was shown $\sqrt{\gamma}T\bar{T}$ is constant along the flow \cite{Guica:2019nzm}, so for the generating functional we have
\begin{align} \label{TTbarGeneratingFunctionalRelation}
 I^{[\lambda]}[\gamma^{[\lambda]}] = I^{[0]}[\gamma^{[0]}] + \frac{\lambda}{2} \int d^2x \sqrt{\gamma}T\bar{T}
\end{align}
where $\sqrt{\gamma}T\bar{T}$ can be taken at any "time" along the flow between $0$ and $\lambda$. In addition, the trace relation also follows from the flow equation \cite{Guica:2019nzm}
\begin{align} \label{TraceRelation}
 T^{[\lambda]} = \frac{c}{24\pi} R[\gamma^{[\lambda]}] - \lambda T\bar{T}^{[\lambda]}
\end{align}
We would like to make a conceptual remark before moving on. There is no physical change of the metric under the $T\bar{T}$ deformation, the flow of the metric is just a mathematical way to describe the flow of the generating functional as a functional of its argument, the metric. The flow equation of the generating functional takes a cleaner form in (\ref{TTbarflowGeneratingFunctionalMetricFixed}). However, the simple relation (\ref{TTbarGeneratingFunctionalRelation}) makes the formulation in terms of a varying metric more convenient for computation, and more importantly, the flow of metric takes compelling physical meaning in the cutoff AdS holography to be discussed later.

\section{Sphere partition function and entanglement entropy}
\setcounter{equation}{0}
\label{sec3}
In this section we compute large $c$ partition function and entanglement entropy of half great circle for $T\bar{T}$ CFT on a sphere of radius $r$ using the $T\bar{T}$ flow equation derived in the previous section. The metric of the sphere is
\begin{align}
 \gamma^{[\lambda]}_{ij} = r^2 \Omega_{ij}
\end{align}
where $\Omega_{ij}$ is the metric of unit sphere. In a maximally symmetric space, the one point function of the energy-momentum tensor is a scalar multiple of the metric
\begin{align}
 T^{[\lambda]}_{ij} = \alpha_\lambda \gamma^{[\lambda]}_{ij}
\end{align}
We can solve $\alpha_\lambda$ from the trace relation as was done in \cite{Donnelly:2018bef}
\begin{align} \label{EMT1ptSphere}
 \alpha_\lambda = \frac{1}{2\lambda} (1-\sqrt{1-\frac{\lambda c}{6\pi r^2}})
\end{align}
and we have $\hat{T}^{[\lambda]}_{ij} = T^{[\lambda]}_{ij} - T^{[\lambda]} \gamma^{[\lambda]}_{ij} = -\alpha_\lambda \gamma^{[\lambda]}_{ij}$. We do $T\bar{T}$ backflow to obtain the metric
\begin{align}
 \gamma^{[0]}_{ij} &= \gamma^{[\lambda]} _{ij}- 2(-\lambda) \hat{T}^{[\lambda]}_{ij} + (-\lambda)^2 \hat{T}^{[\lambda]}_{ik} \gamma^{[\lambda]kl} \hat{T}^{[\lambda]}_{lj} = (1-\lambda \alpha_\lambda)^2 \gamma^{[\lambda]} _{ij}
\end{align}
That is, it's a sphere of radius $r_0 = r(1-\lambda \alpha_\lambda) = \frac{1}{2}(1+\sqrt{1-\frac{\lambda c}{6 \pi r^2}})r$. For the generating functional, we have
\begin{align}
 I^{[\lambda]} &= I^{[0]} + \frac{\lambda}{2} \int d^2x \sqrt{\gamma^{[\lambda]}} {T\bar{T}}^{[\lambda]} = I^{[0]} + \frac{\lambda}{2} 4\pi r^2 (-2\alpha_\lambda^2) \nonumber\\
 &=I^{[0]} + \frac{c}{6} - \frac{2\pi r^2}{\lambda}(1-\sqrt{1-\frac{\lambda c}{6\pi r^2}})
\end{align}
The generating functional of a CFT on a sphere of radius $r_0$ is $I^{[0]} = -\frac{c}{3} \log r_0 + C_0$ where $C_0$ is a constant depending on the UV regularization and renormalization. Plugging in the expression of $r_0$ in terms of $r$, we obtain the generating functional
\begin{align} \label{GeneratingFunctional}
 I^{[\lambda]} = -\log Z^{[\lambda]} = -\frac{c}{3} \log \frac{(1+\sqrt{1-\frac{\lambda c}{6 \pi r^2}})r}{2} + \frac{c}{6} - \frac{2\pi r^2}{\lambda}(1-\sqrt{1-\frac{\lambda c}{6\pi r^2}}) + C_0
\end{align}
One can verify it also satisfies the flow equation with metric fixed (\ref{TTbarflowGeneratingFunctionalMetricFixed}). For comparison with previous work in \cite{Donnelly:2018bef} we switch to the $T\bar{T}$ deformation parameter used there $\mu = -4\lambda$
\begin{align} \label{logZ}
 \log Z^{[\mu]} = \frac{c}{3} \log \frac{(1+\sqrt{1+\frac{\mu c}{24 \pi r^2}})r}{2} - \frac{c}{6} + \frac{8\pi r^2}{\mu}(\sqrt{1+\frac{\mu c}{24\pi r^2}}-1) - C_0
\end{align}
Compared to their result
\begin{align} \label{logZbyDS}
 \log Z^{[\mu]'} = \frac{c}{3} \log (\sqrt{\frac{24\pi}{\mu c}}r + \sqrt{1+\frac{24\pi r^2}{\mu c}}) + \frac{8\pi r^2}{\mu}(\sqrt{1+\frac{\mu c}{24\pi r^2}}-1)
\end{align}
ours differs by a term $\frac{c}{3}\log \frac{\sqrt{\frac{\mu c}{24 \pi}}}{2} - \frac{c}{6} - C_0$ for positive $\mu$ considered in the context in \cite{Donnelly:2018bef}.

For better comparison we briefly review the derivation in \cite{Donnelly:2018bef}. By the definition of energy-momentum tensor we have \footnote{Our definition of energy-momentum tensor takes a different sign than that in \cite{Donnelly:2018bef}, that explains the difference of sign in this equation and others.}
\begin{align}
 r\partial_r \log Z = \int d^2x \sqrt{\gamma} T
\end{align}
The trace of the energy-momentum tensor can be solved from the trace relation as is shown above, which results in a partial differential equation of $\log Z$ of the form
\begin{align}
 r \partial_r \log Z = \frac{16\pi r^2}{\mu}(\sqrt{1+\frac{\mu c}{24 \pi r^2}}-1)
\end{align}
For positive $\mu$, by integration over $r$ we get
\begin{align}
 \log Z = \frac{c}{3} \log (\sqrt{\frac{24\pi}{\mu c}}r + \sqrt{1+\frac{24\pi r^2}{\mu c}}) + \frac{8\pi r^2}{\mu}(\sqrt{1+\frac{\mu c}{24\pi r^2}}-1) + f(\mu)
\end{align}
where $f(\mu)$ is an arbitrary function of $\mu$. By taking the boundary condition $\log Z|_{r=0} = 0$, $f(\mu)$ was set to zero in \cite{Donnelly:2018bef}. We have $f(\mu)=\frac{c}{3}\log \frac{\sqrt{\frac{\mu c}{24 \pi}}}{2} - \frac{c}{6} - C_0$ instead. The partition function (\ref{logZbyDS}) agrees with the simple cutoff AdS holography picture. In that context, the sphere of radius $r$ is embedded in the AdS space as a cutoff surface. In the Schwarzschild coordinates in which the metric takes the form
\begin{align}
 ds^2 = l^2(d\tilde{\rho}^2 + \sinh^2\tilde{\rho} d\Omega_2^2)
\end{align}
the cutoff location is $\tilde{\rho}_c = \sinh^{-1}\frac{r}{l}$. The Brown-York part of the on-shell action of classical gravity 
\begin{align}
 I^{BY} = -\frac{1}{16 \pi G}\int_{\cal B} d^3x \sqrt{g} (R + \frac{2}{l^2}) - \frac{1}{8\pi G} \int_{{\cal M}=\partial{\cal B}} d^2y \sqrt{h} K + \frac{1}{8\pi G} \int_{{\cal M}=\partial{\cal B}}  d^2y \sqrt{h} \frac{1}{l}
\end{align}
agrees with $-\log Z^{[\mu]'}$ by the parameter dictionary (\ref{OldDictionary}) \cite{Caputa:2019pam}\footnote{A sign difference in the Gibbons Hawking term to \cite{Caputa:2019pam} comes from different definition of the extrinsic curvature.}, with the Weyl anomaly counter term $I^{CT}$ neglected.

We think our result (\ref{logZ}) is physically more reasonable in the $\mu \to 0$ limit (CFT limit) compared to (\ref{logZbyDS}). If we start with a CFT with a local Lagrangian and well-defined energy-momentum tensor, we expect perturbative computation of $T\bar{T}$ deformation to work and the partition function should be differentiable at $\mu=0$ to some degree. Our result (\ref{logZ}) is analytic in a neighbourhood of $\mu=0$, in fact by construction, while (\ref{logZbyDS}) is singular at that point. Moreover, (\ref{logZ}) contains the UV regularization term $C_0$ for CFT while (\ref{logZbyDS}) does not. One may be tempted to relate $\mu$ to some effective UV cutoff so (\ref{logZbyDS}) resembles CFT partition function in the $\mu \to 0$ limit, but UV cutoff and $\mu$ are really two different things. We can choose any UV cutoff and renormalization scheme, and $T\bar{T}$ flow with any $\mu$ we want. Another qualitative difference between (\ref{logZ}) and (\ref{logZbyDS}) appears in the $\mu \to \infty$ limit. (\ref{logZbyDS}) goes to zero in accordance with the simple cutoff AdS holography picture, the dimensionless cutoff position $\tilde{\rho}_c$ goes to zero when $\mu$ goes to infinity with $r$ and $c$ fixed. Our result (\ref{logZ}), however, grows without bound as $\mu$ goes to infinity. In the next section we will show that the cutoff AdS holography we derive produces the holographic partition function that agrees with (\ref{logZ}).

By a trick of regarding variation in the replica number $n$ as variation in the metric, it's shown in \cite{Donnelly:2018bef} that the entanglement entropy of a half great circle, with antipodal points being the entangling surface, can be computed from the partition function by the formula
\begin{align}
 S = (1- \frac{r}{2}\partial_r) \log Z
\end{align}
Using the partition function (\ref{logZbyDS}), the entanglement entropy is computed to be
\begin{align}
 S^{[\mu]'} = \frac{c}{3} \log (\sqrt{\frac{24\pi}{\mu c}}r + \sqrt{1+\frac{24\pi r^2}{\mu c}})
\end{align}
which agrees with the geodesic length between the two antipodal points in the cutoff AdS prescribed by the Ryu-Takayanagi formula. It raises the same concern as the partition function though, it doesn't have a well-defined CFT limit. On the other hand, from (\ref{logZ}) the entanglement entropy is computed to be
\begin{align}
 S^{[\mu]} = \frac{c}{3} \log \frac{r(1+\sqrt{1+\frac{\mu c}{24 \pi r^2}})}{2} - \frac{c}{6} - C_0 = \frac{c}{3} \log r_0- \frac{c}{6} - C_0
\end{align}
The entanglement entropy is analytic in a neighbourhood of $\mu=0$ and goes to infinity as $\mu$ goes to infinity.

\section{Cutoff AdS holography}
\setcounter{equation}{0}
\label{sec4}

In this section we show how the cutoff AdS holography works in the sector of classical pure gravity, or on the field theory side, for the large $c$ generating functional without sources other than background metric turned on. We begin with the basic setup of the cutoff AdS holography. We then identify the trace relation of $T\bar{T}$ CFT with the doubly contracted Gauss equation of Brown-York tensor, obtaining a holographic dictionary between $T\bar{T}$ CFT parameters and cutoff AdS gravity parameters. For our key result, we show that the $T\bar{T}$ flow equation of the metric and the one point function of the energy-momentum tensor correspond to flow equation of the boundary metric and Brown-York tensor as the cutoff surface moves in the direction of normal geodesics. We then show the flow equation of the on-shell action of Einstein's gravity in cutoff AdS corresponds to the $T\bar{T}$ flow equation of the generating functional, with straightforward generalization to higher dimensions. We complete this section with a holographic computation of sphere partition function of $T\bar{T}$ CFT.

For a holographic undeformed CFT, it lives on the asymptotic boundary of the dual AdS space. For AdS we use the Fefferman-Graham coordinates in which the metric takes the form
\begin{align}
 ds^2 = \frac{l^2 d\rho^2}{4\rho^2} + g_{ij}(\rho,x) dx^i dx^j
\end{align}
and the asymptotic boundary is located at $\rho=0$. The metric $g_{ij}$ diverges on the asymptotic boundary with a simple pole in $\rho$, it defines a conformal class of the metric of the CFT by
\begin{align}
\gamma_{ij}(x) = \lim_{\rho \to 0}f(\rho,x) g_{ij}(\rho,x)
\end{align}
with $f$ being a arbitrary function with a simple zero at $\rho=0$. Multiplication of a function $e^{2\omega(x)}$ to $f$ corresponds to a Weyl transformation of the metric of the CFT. The canonical choice is $f=\rho$. In the semi-classical limit the generating functional of the CFT is equal to the on-shell action of AdS gravity with prescribed boundary condition on the asymptotic boundary \cite{Gubser:1998bc}\cite{Witten:1998qj}
\begin{align} \label{GKPWRelation}
 I=I_{grav}
\end{align}
with the boundary value of fields $\phi$ corresponding to the sources $J$ in the generating functional of the CFT by the relation 
\begin{align}
J = \lim_{\rho \to 0} \phi \rho^\frac{\Delta-d}{2}
\end{align}
where $\Delta$ is the conformal dimension of the dual operator.

For a  $T\bar{T}$ deformed holographic CFT, it has been proposed to be dual to AdS gravity with cutoff surface at $\rho=\rho_c$ with the boundary metric $h_{ij} = g_{ij}|_{\rho=\rho_c}$. Following the spirit of AdS/CFT, in the semi-classical limit the generating functional of $T\bar{T}$ CFT is equal to the on-shell action of cutoff AdS gravity with prescribed boundary condition on the cutoff surface, with the boundary value of fields $\phi_c$ corresponding to the sources $J$ in the generating functional of $T\bar{T}$ CFT, and the conjugate momenta $\pi_c$ corresponding to the one point functions of the dual operators $O$ \cite{Hartman:2018tkw}
\begin{align} \label{GKPWRelation}
 I[J] &= I_{grav}[\phi_c] \nonumber\\
 J &= \rho_c^\frac{\Delta-d}{2} \phi_c \nonumber\\
 \<O\>_J &= \rho_c^{-\frac{\Delta}{2}} \pi_c
\end{align}
In particular, the boundary metric $h_{ij}$ corresponds to the metric of $T\bar{T}$ CFT $\gamma_{ij}$
\begin{align}
 \gamma_{ij} = \rho_c h_{ij}
\end{align}
and the one point function of the energy-momentum tensor $T_{ij}$ corresponds to the holographic energy-momentum tensor which is Brown-York tensor supplemented by local counter terms 
\begin{align}
 T_{ij} = \rho_c^{1-\frac{d}{2}}{T_{grav}}_{ij} =\rho_c^{1-\frac{d}{2}}(T^{BY}_{ij} + \ldots) = \rho_c^{1-\frac{d}{2}}(-\frac{1}{8\pi G}(K_{ij}-Kh_{ij}+\frac{d-1}{l}h_{ij}) + \ldots)
\end{align}
We have shown the "natural" holographic dictionary for all sources and operators for a would-be complete cutoff AdS holography, however as stated in section \ref{sec2} we restrict ourselves to the $T\bar{T}$ CFT generating functional with other sources turned off, which corresponds to pure AdS gravity. We only expect the cutoff AdS holography to work in this sector \footnote{There have been questions on incorporating matter fields into cutoff AdS holography \cite{Kraus:2018xrn}\cite{Guica:2019nzm}.}. Roughly speaking, the cutoff position $\rho_c$ is related to the $T\bar{T}$ deformation parameter $\lambda$. As $T\bar{T}$ deformation goes the cutoff surface moves inward in the $\rho$ direction, the direction of normal geodesics, with a flow of boundary metric that corresponds to the flow of metric of $T\bar{T}$ CFT shown in (\ref{TTbarflowMetricEMT2dim}).

Now we discuss the cutoff AdS holography for two dimensional $T\bar{T}$ CFT. Starting with the trace relation (\ref{TraceRelation}), we show the holographic dual of the trace relation is doubly contracted Gauss equation for the Brown-York tensor. We have
\begin{align}
 T\bar{T}^{BY} &= (\frac{1}{8\pi G})^2(K_{ij}-Kh_{ij}+\frac{1}{l}h_{ij})(K^{ij}-Kh^{ij}+\frac{1}{l}h^{ij})-(\frac{1}{8\pi G})^2(K-\frac{2}{l})^2 \nonumber\\
 &= (\frac{1}{8\pi G})^2 (K_{ij}K^{ij} - K^2 + \frac{2}{l}K - \frac{2}{l^2})
\end{align}
The Gauss equation, relating intrinsic geometry and extrinsic geometry of the cutoff surface, takes the form \footnote{We refer readers to Appendix B in \cite{Li:2020pwa} or a book on differential geometry like \cite{Spivak:1999}.}
\begin{align} \label{GaussCdntBasis}
\hat{R}_{\rho\sigma\mu\nu} = P^\alpha_\rho P^\beta_\sigma P^\gamma_\mu P^\delta_\nu R_{\alpha\beta\gamma\delta} + K_{\mu\rho}K_{\nu\sigma} - K_{\mu\sigma}K_{\nu\rho}
\end{align}
where $\hat{R}_{\rho\sigma\mu\nu}$ is the curvature tensor of intrinsic geometry of the cutoff surface, $R_{\alpha\beta\gamma\delta}$ is the curvature tensor of the ambient space and $P^\alpha_\rho$ is the projection operator to the tangent space of the cutoff surface. We used Greek indices to denote coordinates in the ambient space. In the Gaussian normal coordinates, we use $\rm s$ to denote the arclength coordinate of normal geodesics with the unit normal being $n=\partial_{\rm s}$, and Roman indices to denote the transverse coordinates.
By a double contraction of the Gauss equation we get
\begin{align}
\hat{R} = R - 2 R_{\alpha\beta}n^\alpha n^\beta + K^2 - K_{ij}K^{ij}
\end{align}
For pure gravity $R_{\alpha\beta} = -\frac{2}{l^2}g_{\alpha\beta}$ and $R=-\frac{6}{l^2}$, so we have $K_{ij}K^{ij} - K^2 = -\hat{R} - \frac{2}{l^2}$. Plugging in this expression, we obtain
\begin{align}
 T\bar{T}^{BY} = (\frac{1}{8\pi G})^2 (-\hat{R} + \frac{2}{l}K - \frac{4}{l^2}) = -(\frac{1}{8\pi G})^2\hat{R} + \frac{1}{4\pi Gl}T^{BY}
\end{align}
or
\begin{align}
 T^{BY} = 4\pi G l T\bar{T}^{BY}  + \frac{l}{16\pi G} \hat{R}
\end{align}
which holographically translates to
\begin{align}
 \rho_c T = 4\pi G l \rho_c^2 T\bar{T} + \frac{l}{16\pi G}\rho_c R(\gamma)
\end{align}
we see it agrees with the trace relation (\ref{TraceRelation}) with the holographic dictionary
\begin{align}
 c=\frac{3l}{2G}, \quad \lambda=-4\pi Gl\rho_c
\end{align}
or
\begin{align}
 c=\frac{3l}{2G}, \quad \mu=16\pi Gl\rho_c
\end{align}
This has been obtained in \cite{Caputa:2020lpa}.

An important point of \cite{Guica:2019nzm} is the solution of the metric to the $T\bar{T}$ flow equation (\ref{TTbarflowSol2dim}) holographically corresponds to a mixed boundary condition for graviton, which happens to be a Dirichlet boundary condition on a cutoff surface due to special properties of three dimensional gravity, that is, the Fefferman-Graham expansion of the metric terminates. Similar approach to holographic dual of $T\bar{T}$ CFT using the Chern-Simons formulation of three dimensional gravity was studied in \cite{Llabres:2019jtx}. However, we think the flow equation (\ref{TTbarflowMetricEMT2dim}) is more fundamental to the cutoff AdS picture than the form of the solution because it holographically corresponds to the flow equation of the boundary metric and Brown-York tensor as the cutoff surface moves along the direction of normal geodesics, and this notion doesn't depend on the dimension of the space nor the choice of the Fefferman-Graham gauge. With the holographic dictionary we find
\begin{align}
 \partial_\lambda \gamma_{ij} = -\frac{1}{4\pi Gl} \partial_\rho (\rho h_{ij}) = -\frac{1}{4\pi Gl} h_{ij} + \frac{1}{8\pi G} \partial_n h_{ij} = \frac{1}{4\pi G}(K_{ij} - \frac{1}{l}h_{ij}) = -2{\hat{T}^{BY}}_{ij} = -2\hat{T}_{ij}
\end{align}
and the other equation corresponds to
\begin{align} \label{1ptFlowHolographicDual}
 &-\frac{\rho}{4\pi Gl}\partial_\rho(-\frac{1}{8\pi G}(K_{ij}-\frac{1}{l}h_{ij})) = \frac{1}{2} (\frac{1}{8\pi G})^2 (K_{ij}K^{ij} - K^2 + \frac{2}{l}K - \frac{2}{l^2})h_{ij} \nonumber\\
 &-2 (\frac{1}{8\pi G})^2 (K_{ik}K_j^k - \frac{2}{l}K_{ij} + \frac{1}{l^2}h_{ij}) + (\frac{1}{8\pi G})^2 (KK_{ij}-\frac{1}{l}Kh_{ij} - \frac{2}{l}K_{ij} + \frac{2}{l^2}h_{ij})
\end{align}
We have
\begin{align}
 -2\frac{\rho}{l}\partial_\rho K_{ij} = \partial_{\rm s} K_{ij} &= \nabla_{\rm s} (g(\nabla_i \partial_{\rm s},\partial_j)) = g(\nabla_{\rm s}\nabla_i \partial_{\rm s},\partial_j) + g(\nabla_i \partial_{\rm s},\nabla_{\rm s} \partial_j) \nonumber\\
 &= g((\nabla_{\rm s}\nabla_i-\nabla_i\nabla_{\rm s}) \partial_{\rm s},\partial_j) + g(\nabla_i \partial_{\rm s},\nabla_j \partial_{\rm s}) \nonumber\\
 &= R_{jssi} + K_{ik}K_j^k
\end{align}
where we used $\nabla_{\rm s}\partial_{\rm s}=0$ by the definition of normal geodesics, $\nabla_{\rm s}\partial_j = \nabla_j \partial_{\rm s}$ by torsion-freeness of the covariant derivative and $\nabla_{\rm s}g = 0$ by metric compatibility of the covariant derivative. With the singly contracted Gauss equation
\begin{align}
 \hat{R}_{ij} = R_{ij} - R_{sisj} + K K_{ij} - K_{ik} K_j^k
\end{align}
(\ref{1ptFlowHolographicDual}) reduces to
\begin{align}
 \hat{R}_{ij} - \frac{1}{2}\hat{R} h_{ij} = 0
\end{align}
which holds identically in two dimensional space.

The holographic interpretation of $T\bar{T}$ flow equation given above can be generalized to a higher dimension $d$, with the generalization of $T\bar{T}$ operator
\begin{align}
 T\bar{T} = T_{ij}T^{ij} - \frac{1}{d-1}T^2
\end{align}
up to local counter terms. In fact, the most direct way to see the cutoff AdS holography for $T\bar{T}$ CFT is by flow equation of the gravity on-shell action as the cutoff surface moves in the direction of normal geodesics. For the Brown-York part of the gravity on-shell action, the flow equation is simply
\begin{align}
 \partial_n I^{BY} = \frac{1}{2} \int \sqrt{h} d^dx T^{BY}_{ij} {\cal L}_n h^{ij}
\end{align}
where $\cal L$ denotes Lie derivative. We have ${\cal L}_n h^{ij} = -2K^{ij} = -2(-8\pi G {\hat{T}^{BYij}}+\frac{1}{l} h^{ij})$ where we define in $d$ dimension $\hat{T}^{BYij} = T^{BYij} - \frac{1}{d-1} T^{BY} h^{ij}$ and $T\bar{T}^{BY} = T^{BY}_{ij}T^{BYij} - \frac{1}{d-1} (T^{BY})^2 = \hat{T}^{BY}_{ij}\hat{T}^{BYij} - (\hat{T}^{BY})^2 = T^{BY}_{ij} \hat{T}^{BYij}$. Now we get
\begin{align}
 \partial_n I^{BY} = 8\pi G \int d^dx \sqrt{h} T\bar{T}^{BY} - \int d^dx \sqrt{h} \frac{1}{l} T^{BY}
\end{align}
Using $T^{BY} = 4\pi G l T\bar{T}^{BY} + \frac{l}{16\pi G}\hat{R}$ we obtain
\begin{align}
 \partial_n I^{BY} = -2\frac{\rho}{l}\partial_\rho I^{BY} =  4\pi G \int d^dx \sqrt{h} T\bar{T}^{BY} - \frac{1}{16\pi G} \int d^dx \sqrt{h} \hat{R}
\end{align}
We see the flow equation of the gravity on-shell action takes the form of $T\bar{T}$ flow up to local counter terms, with the $T\bar{T}$ deformation parameter related to the AdS cutoff position by
\begin{align}
 \lambda = -\frac{8\pi Gl}{d}\rho_c^{\frac{d}{2}}
\end{align}
Now we specialize to dimension two, in which $\int d^2x \sqrt{h} \hat{R}$ is a topological invariant $4\pi \chi(\cal M)$. So we have
\begin{align}
 \partial_\rho (I^{BY} - \frac{l}{32\pi G} \log\rho \int d^2x \sqrt{h} \hat{R}) = -\frac{2\pi G l}{\rho} \int d^2 \sqrt{h} T\bar{T}^{BY}
\end{align}
But $I^{CT} = - \frac{l}{32\pi G} \log\rho \int d^2x \sqrt{h} \hat{R}$ is just the Weyl anomaly counter term in dimension two \cite{Henningson:1998gx}\cite{deHaro:2000vlm}, we simply have
\begin{align}
 \partial_\rho I_{grav} = -\frac{2\pi G l}{\rho} \int d^2x \sqrt{h} T\bar{T}_{grav}
\end{align}
and the holographic dual equation
\begin{align}
 \partial_\lambda I = \frac{1}{2} \int d^2x \sqrt{\gamma} T\bar{T}
\end{align}
on the field theory side. This is an infinitesimal version of the observation in \cite{Caputa:2020lpa} that the gravity on-shell action in the annular region between two cutoff surfaces is equal to integration of $T\bar{T}$ on either of the two cutoff surfaces, multiplied by the difference in the $T\bar{T}$ deformation parameter.

As an example of the cutoff AdS holography, we complete this section by a computation of the holographic partition function of $T\bar{T}$ CFT on a sphere of radius $r$. The metric of $\rm AdS_3$ takes the form in the Schwarzschild coordinates
\begin{align}
 ds^2 = l^2(d\tilde{\rho}^2 + \sinh^2 \tilde{\rho} d\Omega_2^2 )
\end{align}
or in the Fefferman-Graham coordinates
\begin{align}
 ds^2 = \frac{l^2 d\rho^2}{4\rho^2} + (\frac{r_0^2}{\rho}-\frac{l^2}{2}+\frac{\rho l^4}{16 r_0^2}) d\Omega_2^2
\end{align}
with $\rho=\frac{4 r_0^2}{l^2} e^{-2\tilde{\rho}}$. The mathematically related undeformed CFT lives on the asymptotic boundary with metric $r_0^2 d\Omega_2^2$ while the boundary metric at the cutoff position $\rho_c$ corresponds to the metric of the $T\bar{T}$ CFT
\begin{align}
 \rho_c h_{ij}(\rho_c) dx^i dx^j = r_0^2 (1-\frac{\rho_c l^2}{4r_0^2})^2 d\Omega_2^2 = r^2 d\Omega_2^2
\end{align}
With the holographic dictionary $\lambda = -4\pi Gl \rho_c$, we can solve the cutoff position $\rho_c$ and $r_0$ in terms of $r$ and $\lambda$
\begin{align}
 & \rho_c = -\frac{\lambda}{4\pi Gl} \nonumber\\
 & r_0 = \frac{1+ \sqrt{1-\frac{\lambda l}{4\pi G r^2}}}{2} r
\end{align}
It's straightforward to compute the Brown-York part of the on-shell action with the cutoff at $\rho_c$
\begin{align}
 I^{BY} = \frac{l}{4G}\log\frac{\rho_c l^2}{4r_0^2} + \frac{l}{4G}(\frac{\rho_c l^2}{4r_0^2} -1)
\end{align}
and the Weyl anomaly counter term
\begin{align}
 I^{CT} = - \frac{l}{32\pi G} \log \rho_c \int d^2x \sqrt{h} \hat{R} = - \frac{l}{32\pi G} \log \rho_c 4\pi \chi({\cal M}) = -\frac{l}{4G} \log \rho_c
\end{align}
We add up to obtain
\begin{align}
 I_{grav} = I^{BY} + I^{CT} &=  - \frac{l}{2G} \log r_0 + \frac{\rho_c l^3}{16G r_0^2} - \frac{l}{4G}(1+\log\frac{4}{l^2}) \nonumber\\
 &= -\frac{l}{2G} \log (\frac{1+\sqrt{1-\frac{\lambda c}{4\pi G r^2}}}{2}r) - \frac{2\pi r^2}{\lambda}(1-\sqrt{1-\frac{\lambda c}{4\pi G r^2}}) - \frac{l}{4G} \log\frac{4}{l^2}
\end{align}
We see it agrees with the $T\bar{T}$ CFT generating functional (\ref{GeneratingFunctional}) with the central charge relation $c=\frac{3l}{2G}$.

\section{Summary and discussion}
\setcounter{equation}{0}
\label{sec5}
In this article we obtained large $c$ sphere partition function and entanglement entropy of half great circle for $T\bar{T}$ deformed two dimensional conformal field theory with reasonable CFT limits, using the $T\bar{T}$ flow equation derived in \cite{Guica:2019nzm}. We found the partition function and entanglement entropy grow without bound with the $T\bar{T}$ deformation parameter $\mu$, roughly as $\log Z \sim \log\mu$. We hope to get better understanding of this behavior, in particular its relation to non-locality of the theory. For cutoff AdS holography we restricted ourselves to the sector of classical pure gravity, and showed that the $T\bar{T}$ flow equations of the metric, the one point function of the energy-momentum tensor and the generating functional correspond to the flow equations of the holographic dual quantities as the cutoff surface moves inward, with straightforward generalization to higher dimensions. We do not expect the cutoff AdS holography to work as a full holography with matter fields included, as is indicated by the large $\mu$ behavior of sphere partition function that it doesn't go to zero. In addition, a full cutoff AdS holography would go against the integrability of the $T\bar{T}$ deformation, perturbative computation of correlators and other considerations \cite{Kraus:2018xrn}\cite{Guica:2019nzm}. Perhaps it's possible to add double trace operators dual to matter fields to $T\bar{T}$ deformation, with fine-tuned coefficients, to move the matter fields into the bulk together with gravity. But it may be better to take a broader perspective and think about what a holography in finite space could be with the $T\bar{T}$ as a reference point in the sector of classical pure gravity.


\section*{Acknowledgements}

We would like to thank Per Kraus for his thoughtful comments on the note. We also thank Yang Zhou, Song He and Ruben Monten for discussion.

\newpage

\appendix

\newpage


\begin{thebibliography}{99}
\small

\bibitem{Zamolodchikov:2004ce}
  A.~B.~Zamolodchikov,
  hep-th/0401146.

\bibitem{Smirnov:2016lqw}
  F.~A.~Smirnov and A.~B.~Zamolodchikov,
  Nucl.\ Phys.\ B {\bf 915} (2017) 363
  doi:10.1016/j.nuclphysb.2016.12.014
  [arXiv:1608.05499 [hep-th]].
 
\bibitem{Cavaglia:2016oda}
  A.~Cavaglià, S.~Negro, I.~M.~Szécsényi and R.~Tateo,
  JHEP {\bf 1610} (2016) 112
  doi:10.1007/JHEP10(2016)112
  [arXiv:1608.05534 [hep-th]].
  
\bibitem{Dubovsky:2012wk}
S.~Dubovsky, R.~Flauger and V.~Gorbenko,
JHEP \textbf{09} (2012), 133
doi:10.1007/JHEP09(2012)133
[arXiv:1205.6805 [hep-th]].
  
\bibitem{Dubovsky:2017cnj}
  S.~Dubovsky, V.~Gorbenko and M.~Mirbabayi,
  JHEP {\bf 1709} (2017) 136
  doi:10.1007/JHEP09(2017)136
  [arXiv:1706.06604 [hep-th]].
  
\bibitem{Dubovsky:2018bmo}
  S.~Dubovsky, V.~Gorbenko and G.~Hernández-Chifflet,
  JHEP {\bf 1809} (2018) 158
  doi:10.1007/JHEP09(2018)158
  [arXiv:1805.07386 [hep-th]].
  
\bibitem{Frolov:2019nrr}
S.~Frolov,
Proc. Steklov Inst. Math. \textbf{309} (2020), 107-126
doi:10.1134/S0081543820030098
[arXiv:1905.07946 [hep-th]].

\bibitem{Sfondrini:2019smd}
A.~Sfondrini and S.~J.~van Tongeren,
Phys. Rev. D \textbf{101} (2020) no.6, 066022
doi:10.1103/PhysRevD.101.066022
[arXiv:1908.09299 [hep-th]].
  
\bibitem{Callebaut:2019omt}
N.~Callebaut, J.~Kruthoff and H.~Verlinde,
JHEP \textbf{04} (2020), 084
doi:10.1007/JHEP04(2020)084
[arXiv:1910.13578 [hep-th]].

\bibitem{Jorjadze:2020ili}
G.~Jorjadze and S.~Theisen,
[arXiv:2001.03563 [hep-th]].
  
\bibitem{Conti:2018jho}
  R.~Conti, L.~Iannella, S.~Negro and R.~Tateo,
  JHEP {\bf 1811} (2018) 007
  doi:10.1007/JHEP11(2018)007
  [arXiv:1806.11515 [hep-th]].
  
\bibitem{LeFloch:2019wlf}
B.~Le Floch and M.~Mezei,
SciPost Phys. \textbf{7} (2019) no.4, 043
doi:10.21468/SciPostPhys.7.4.043
[arXiv:1907.02516 [hep-th]].

\bibitem{Chen:2021aid}
B.~Chen, J.~Hou and J.~Tian,
[arXiv:2102.01470 [hep-th]].

\bibitem{Guica:2020uhm}
M.~Guica and R.~Monten,
[arXiv:2011.05445 [hep-th]].

\bibitem{Guica:2020eab}
M.~Guica,
[arXiv:2012.15806 [hep-th]].
  
\bibitem{Datta:2018thy}
  S.~Datta and Y.~Jiang,
  JHEP {\bf 1808} (2018) 106
  doi:10.1007/JHEP08(2018)106
  [arXiv:1806.07426 [hep-th]].
  
\bibitem{Aharony:2018bad}
  O.~Aharony, S.~Datta, A.~Giveon, Y.~Jiang and D.~Kutasov,
  JHEP {\bf 1901} (2019) 086
  doi:10.1007/JHEP01(2019)086
  [arXiv:1808.02492 [hep-th]].

\bibitem{Bonelli:2018kik}
G.~Bonelli, N.~Doroud and M.~Zhu,
JHEP \textbf{06} (2018), 149
doi:10.1007/JHEP06(2018)149
[arXiv:1804.10967 [hep-th]].
  
\bibitem{Rosenhaus:2019utc}
V.~Rosenhaus and M.~Smolkin,
Phys. Rev. D \textbf{102} (2020) no.6, 065009
doi:10.1103/PhysRevD.102.065009
[arXiv:1909.02640 [hep-th]].
  
\bibitem{Cardy:2018sdv}
  J.~Cardy,
  JHEP {\bf 1810} (2018) 186
  doi:10.1007/JHEP10(2018)186
  [arXiv:1801.06895 [hep-th]].
  
\bibitem{Donnelly:2018bef}
  W.~Donnelly and V.~Shyam,
  Phys.\ Rev.\ Lett.\  {\bf 121} (2018) no.13,  131602
  doi:10.1103/PhysRevLett.121.131602
  [arXiv:1806.07444 [hep-th]].
  
\bibitem{Santilli:2018xux}
L.~Santilli and M.~Tierz,
JHEP \textbf{01} (2019), 054
doi:10.1007/JHEP01(2019)054
[arXiv:1810.05404 [hep-th]].
  
\bibitem{Caputa:2019pam}
  P.~Caputa, S.~Datta and V.~Shyam,
  JHEP {\bf 1905} (2019) 112
  doi:10.1007/JHEP05(2019)112
  [arXiv:1902.10893 [hep-th]].
  
\bibitem{He:2020cxp}
S.~He, Y.~Sun and Y.~X.~Zhang,
[arXiv:2011.02902 [hep-th]].

\bibitem{Kraus:2018xrn}
  P.~Kraus, J.~Liu and D.~Marolf,
  JHEP {\bf 1807} (2018) 027
  doi:10.1007/JHEP07(2018)027
  [arXiv:1801.02714 [hep-th]].

\bibitem{Aharony:2018vux}
  O.~Aharony and T.~Vaknin,
  JHEP {\bf 1805} (2018) 166
  doi:10.1007/JHEP05(2018)166
  [arXiv:1803.00100 [hep-th]].
  
\bibitem{Cardy:2019qao}
J.~Cardy,
JHEP \textbf{19} (2020), 160
doi:10.1007/JHEP12(2019)160
[arXiv:1907.03394 [hep-th]].

\bibitem{He:2019vzf}
S.~He and H.~Shu,
JHEP \textbf{02} (2020), 088
doi:10.1007/JHEP02(2020)088
[arXiv:1907.12603 [hep-th]].

\bibitem{He:2019ahx}
S.~He, J.~R.~Sun and Y.~Sun,
JHEP \textbf{04} (2020), 100
doi:10.1007/JHEP04(2020)100
[arXiv:1912.11461 [hep-th]].

\bibitem{He:2020udl}
S.~He and Y.~Sun,
Phys. Rev. D \textbf{102} (2020) no.2, 026023
doi:10.1103/PhysRevD.102.026023
[arXiv:2004.07486 [hep-th]].

\bibitem{Li:2020pwa}
Y.~Li and Y.~Zhou,
JHEP \textbf{12} (2020), 168
doi:10.1007/JHEP12(2020)168
[arXiv:2005.01693 [hep-th]].

\bibitem{Hirano:2020ppu}
S.~Hirano, T.~Nakajima and M.~Shigemori,
[arXiv:2012.03972 [hep-th]].

\bibitem{He:2020qcs}
S.~He,
[arXiv:2012.06202 [hep-th]].

\bibitem{Baggio:2018rpv}
M.~Baggio, A.~Sfondrini, G.~Tartaglino-Mazzucchelli and H.~Walsh,
JHEP \textbf{06} (2019), 063
doi:10.1007/JHEP06(2019)063
[arXiv:1811.00533 [hep-th]].

\bibitem{Chang:2018dge}
C.~K.~Chang, C.~Ferko and S.~Sethi,
JHEP \textbf{04} (2019), 131
doi:10.1007/JHEP04(2019)131
[arXiv:1811.01895 [hep-th]].

\bibitem{Jiang:2019hux}
H.~Jiang, A.~Sfondrini and G.~Tartaglino-Mazzucchelli,
Phys. Rev. D \textbf{100} (2019) no.4, 046017
doi:10.1103/PhysRevD.100.046017
[arXiv:1904.04760 [hep-th]].

\bibitem{Chang:2019kiu}
C.~K.~Chang, C.~Ferko, S.~Sethi, A.~Sfondrini and G.~Tartaglino-Mazzucchelli,
Phys. Rev. D \textbf{101} (2020) no.2, 026008
doi:10.1103/PhysRevD.101.026008
[arXiv:1906.00467 [hep-th]].
  
\bibitem{Jiang:2019tcq}
  Y.~Jiang,
  JHEP {\bf 2002} (2020) 094
  doi:10.1007/JHEP02(2020)094
  [arXiv:1903.07561 [hep-th]].
  
\bibitem{Brennan:2020dkw}
T.~D.~Brennan, C.~Ferko, E.~Martinec and S.~Sethi,
[arXiv:2005.00431 [hep-th]].
  
\bibitem{Tolley:2019nmm}
A.~J.~Tolley,
[arXiv:1911.06142 [hep-th]].

\bibitem{Mazenc:2019cfg}
E.~A.~Mazenc, V.~Shyam and R.~M.~Soni,
[arXiv:1912.09179 [hep-th]].


\bibitem{Chakraborty:2018kpr}
S.~Chakraborty, A.~Giveon, N.~Itzhaki and D.~Kutasov,
Nucl. Phys. B \textbf{935} (2018), 290-309
doi:10.1016/j.nuclphysb.2018.08.011
[arXiv:1805.06286 [hep-th]].

\bibitem{Chen:2018eqk}
  B.~Chen, L.~Chen and P.~X.~Hao,
  Phys.\ Rev.\ D {\bf 98} (2018) no.8,  086025
  doi:10.1103/PhysRevD.98.086025
  [arXiv:1807.08293 [hep-th]].
  
\bibitem{Banerjee:2019ewu}
  A.~Banerjee, A.~Bhattacharyya and S.~Chakraborty,
  Nucl.\ Phys.\ B {\bf 948} (2019) 114775
  doi:10.1016/j.nuclphysb.2019.114775
  [arXiv:1904.00716 [hep-th]].
  
\bibitem{Murdia:2019fax}
C.~Murdia, Y.~Nomura, P.~Rath and N.~Salzetta,
Phys. Rev. D \textbf{100} (2019) no.2, 026011
doi:10.1103/PhysRevD.100.026011
[arXiv:1904.04408 [hep-th]].

\bibitem{Ota:2019yfe}
T.~Ota,
[arXiv:1904.06930 [hep-th]].
  
\bibitem{Jeong:2019ylz}
H.~Jeong, K.~Kim and M.~Nishida,
Phys. Rev. D \textbf{100} (2019) no.10, 106015
doi:10.1103/PhysRevD.100.106015
[arXiv:1906.03894 [hep-th]].

\bibitem{Grieninger:2019zts}
S.~Grieninger,
JHEP \textbf{11} (2019), 171
doi:10.1007/JHEP11(2019)171
[arXiv:1908.10372 [hep-th]].

\bibitem{Lewkowycz:2019xse}
A.~Lewkowycz, J.~Liu, E.~Silverstein and G.~Torroba,
JHEP \textbf{04} (2020), 152
doi:10.1007/JHEP04(2020)152
[arXiv:1909.13808 [hep-th]].

\bibitem{Geng:2019ruz}
H.~Geng,
JHEP \textbf{02} (2020), 005
doi:10.1007/JHEP02(2020)005
[arXiv:1911.02644 [hep-th]].

\bibitem{Donnelly:2019pie}
W.~Donnelly, E.~LePage, Y.~Li, A.~Pereira and V.~Shyam,
[arXiv:1909.11402 [hep-th]].

\bibitem{Chakraborty:2020udr}
S.~Chakraborty and A.~Hashimoto,
[arXiv:2010.15759 [hep-th]].

\bibitem{Khoeini-Moghaddam:2020ymm}
S.~Khoeini-Moghaddam, F.~Omidi and C.~Paul,
[arXiv:2011.00305 [hep-th]].



\bibitem{Conti:2018tca}
R.~Conti, S.~Negro and R.~Tateo,
JHEP \textbf{02} (2019), 085
doi:10.1007/JHEP02(2019)085
[arXiv:1809.09593 [hep-th]].

\bibitem{Jafari:2019qns}
G.~Jafari, A.~Naseh and H.~Zolfi,
Phys. Rev. D \textbf{101} (2020) no.2, 026007
doi:10.1103/PhysRevD.101.026007
[arXiv:1909.02357 [hep-th]].

\bibitem{Gross:2019ach}
D.~J.~Gross, J.~Kruthoff, A.~Rolph and E.~Shaghoulian,
Phys. Rev. D \textbf{101} (2020) no.2, 026011
doi:10.1103/PhysRevD.101.026011
[arXiv:1907.04873 [hep-th]].

\bibitem{Ireland:2019vvj}
A.~Ireland and V.~Shyam,
JHEP \textbf{07} (2020), 058
doi:10.1007/JHEP07(2020)058
[arXiv:1912.04686 [hep-th]].

\bibitem{Haruna:2020wjw}
J.~Haruna, T.~Ishii, H.~Kawai, K.~Sakai and K.~Yoshida,
JHEP \textbf{04} (2020), 127
doi:10.1007/JHEP04(2020)127
[arXiv:2002.01414 [hep-th]].

\bibitem{Leoni:2020rof}
M.~Leoni,
JHEP \textbf{07} (2020) no.07, 230
doi:10.1007/JHEP07(2020)230
[arXiv:2005.08906 [hep-th]].

\bibitem{Kruthoff:2020hsi}
J.~Kruthoff and O.~Parrikar,
[arXiv:2006.03054 [hep-th]].

\bibitem{Ouyang:2020rpq}
H.~Ouyang and H.~Shu,
Eur. Phys. J. C \textbf{80} (2020) no.12, 1155
doi:10.1140/epjc/s10052-020-08738-6
[arXiv:2006.10514 [hep-th]].

\bibitem{Santilli:2020qvd}
L.~Santilli, R.~J.~Szabo and M.~Tierz,
JHEP \textbf{11} (2020), 086
doi:10.1007/JHEP11(2020)086
[arXiv:2009.00657 [hep-th]].

\bibitem{Jiang:2020nnb}
Y.~Jiang,
[arXiv:2011.00637 [hep-th]].

\bibitem{He:2020hhm}
M.~He and Y.~h.~Gao,
[arXiv:2012.05726 [hep-th]].

\bibitem{Hansen:2020hrs}
D.~Hansen, Y.~Jiang and J.~Xu,
[arXiv:2012.12290 [hep-th]].


\bibitem{Giveon:2017myj}
A.~Giveon, N.~Itzhaki and D.~Kutasov,
JHEP \textbf{12} (2017), 155
doi:10.1007/JHEP12(2017)155
[arXiv:1707.05800 [hep-th]].

\bibitem{Hirano:2020nwq}
S.~Hirano and M.~Shigemori,
JHEP \textbf{11} (2020), 108
doi:10.1007/JHEP11(2020)108
[arXiv:2003.06300 [hep-th]].

\bibitem{McGough:2016lol}
  L.~McGough, M.~Mezei and H.~Verlinde,
  JHEP {\bf 1804} (2018) 010
  doi:10.1007/JHEP04(2018)010
  [arXiv:1611.03470 [hep-th]].
  
\bibitem{Balasubramanian:1999re}
V.~Balasubramanian and P.~Kraus,
Commun. Math. Phys. \textbf{208} (1999), 413-428
doi:10.1007/s002200050764
[arXiv:hep-th/9902121 [hep-th]].

\bibitem{Henningson:1998gx}
  M.~Henningson and K.~Skenderis,
  JHEP {\bf 9807} (1998) 023
  doi:10.1088/1126-6708/1998/07/023
  [hep-th/9806087].
  
\bibitem{deHaro:2000vlm}
  S.~de Haro, S.~N.~Solodukhin and K.~Skenderis,
  Commun.\ Math.\ Phys.\  {\bf 217} (2001) 595
  doi:10.1007/s002200100381
  [hep-th/0002230].
  
\bibitem{Taylor:2018xcy}
  M.~Taylor,
  arXiv:1805.10287 [hep-th].

\bibitem{Ryu:2006bv}
S.~Ryu and T.~Takayanagi,
Phys. Rev. Lett. \textbf{96} (2006), 181602
doi:10.1103/PhysRevLett.96.181602
[arXiv:hep-th/0603001 [hep-th]].

\bibitem{Lewkowycz:2013nqa}
A.~Lewkowycz and J.~Maldacena,
JHEP \textbf{08} (2013), 090
doi:10.1007/JHEP08(2013)090
[arXiv:1304.4926 [hep-th]].
  
\bibitem{Hartman:2018tkw}
  T.~Hartman, J.~Kruthoff, E.~Shaghoulian and A.~Tajdini,
  JHEP {\bf 1903} (2019) 004
  doi:10.1007/JHEP03(2019)004
  [arXiv:1807.11401 [hep-th]].
  
\bibitem{Caputa:2020lpa}
P.~Caputa, S.~Datta, Y.~Jiang and P.~Kraus,
[arXiv:2011.04664 [hep-th]].

\bibitem{Iliesiu:2020zld}
L.~V.~Iliesiu, J.~Kruthoff, G.~J.~Turiaci and H.~Verlinde,
[arXiv:2004.07242 [hep-th]].

\bibitem{Belin:2020oib}
A.~Belin, A.~Lewkowycz and G.~Sarosi,
JHEP \textbf{09} (2020), 156
doi:10.1007/JHEP09(2020)156
[arXiv:2006.01835 [hep-th]].

\bibitem{Guica:2019nzm}
M.~Guica and R.~Monten,
SciPost Phys. \textbf{10} (2021) no.2, 024
doi:10.21468/SciPostPhys.10.2.024
[arXiv:1906.11251 [hep-th]].
  
  
\bibitem{Witten:2001ua}
E.~Witten,
[arXiv:hep-th/0112258 [hep-th]].

\bibitem{Berkooz:2002ug}
M.~Berkooz, A.~Sever and A.~Shomer,
JHEP \textbf{05} (2002), 034
doi:10.1088/1126-6708/2002/05/034
[arXiv:hep-th/0112264 [hep-th]].

\bibitem{Mueck:2002gm}
W.~Mueck,
Phys. Lett. B \textbf{531} (2002), 301-304
doi:10.1016/S0370-2693(02)01487-9
[arXiv:hep-th/0201100 [hep-th]].

\bibitem{Minces:2002wp}
P.~Minces,
Phys. Rev. D \textbf{68} (2003), 024027
doi:10.1103/PhysRevD.68.024027
[arXiv:hep-th/0201172 [hep-th]].

\bibitem{Sever:2002fk}
A.~Sever and A.~Shomer,
JHEP \textbf{07} (2002), 027
doi:10.1088/1126-6708/2002/07/027
[arXiv:hep-th/0203168 [hep-th]].

\bibitem{Gubser:2002zh}
S.~S.~Gubser and I.~Mitra,
Phys. Rev. D \textbf{67} (2003), 064018
doi:10.1103/PhysRevD.67.064018
[arXiv:hep-th/0210093 [hep-th]].

\bibitem{Gubser:2002vv}
S.~S.~Gubser and I.~R.~Klebanov,
Nucl. Phys. B \textbf{656} (2003), 23-36
doi:10.1016/S0550-3213(03)00056-7
[arXiv:hep-th/0212138 [hep-th]].

\bibitem{Aharony:2005sh}
O.~Aharony, M.~Berkooz and B.~Katz,
JHEP \textbf{10} (2005), 097
doi:10.1088/1126-6708/2005/10/097
[arXiv:hep-th/0504177 [hep-th]].

\bibitem{Elitzur:2005kz}
S.~Elitzur, A.~Giveon, M.~Porrati and E.~Rabinovici,
JHEP \textbf{02} (2006), 006
doi:10.1088/1126-6708/2006/02/006
[arXiv:hep-th/0511061 [hep-th]].

\bibitem{Hartman:2006dy}
T.~Hartman and L.~Rastelli,
JHEP \textbf{01} (2008), 019
doi:10.1088/1126-6708/2008/01/019
[arXiv:hep-th/0602106 [hep-th]].

\bibitem{Diaz:2007an}
D.~E.~Diaz and H.~Dorn,
JHEP \textbf{05} (2007), 046
doi:10.1088/1126-6708/2007/05/046
[arXiv:hep-th/0702163 [hep-th]].
  
\bibitem{Papadimitriou:2007sj}
I.~Papadimitriou,
JHEP \textbf{05} (2007), 075
doi:10.1088/1126-6708/2007/05/075
[arXiv:hep-th/0703152 [hep-th]].

\bibitem{Giombi:2018vtc}
S.~Giombi, V.~Kirilin and E.~Perlmutter,
JHEP \textbf{02} (2018), 175
doi:10.1007/JHEP02(2018)175
[arXiv:1801.01477 [hep-th]].


\bibitem{Bzowski:2018pcy}
A.~Bzowski and M.~Guica,
JHEP \textbf{01} (2019), 198
doi:10.1007/JHEP01(2019)198
[arXiv:1803.09753 [hep-th]].

\bibitem{Gubser:1998bc}
S.~S.~Gubser, I.~R.~Klebanov and A.~M.~Polyakov,
Phys. Lett. B \textbf{428} (1998), 105-114
doi:10.1016/S0370-2693(98)00377-3
[arXiv:hep-th/9802109 [hep-th]].

\bibitem{Witten:1998qj}
E.~Witten,
Adv. Theor. Math. Phys. \textbf{2} (1998), 253-291
doi:10.4310/ATMP.1998.v2.n2.a2
[arXiv:hep-th/9802150 [hep-th]].

\bibitem{Spivak:1999}
M.~Spivak, ``A Comprehensive Introduction to Differential Geometry", Volume 3, Publish or Perish, Inc., 1999
  
\bibitem{Llabres:2019jtx}
E.~Llabr\'es,
[arXiv:1912.13330 [hep-th]].
  
\end{thebibliography}
\end{document}